\documentclass[usenatbib]{mn2e}
\usepackage{amssymb}
\usepackage{color}
\usepackage{graphicx}

\newcommand{\be}{\begin{equation}}
\newcommand{\ee}{\end{equation}}
\newcommand{\ba}{\begin{eqnarray}}
\newcommand{\ea}{\end{eqnarray}}
\newcommand{\lp}{\left(}
\newcommand{\rp}{\right)}

\newcommand{\Pdot}{\dot{P}}

\title[Magnetic field growth in young pulsars]{Magnetic field growth
in young glitching pulsars with a braking index}
\author[Ho]{Wynn C. G. Ho,$^{1}$\thanks{Email: wynnho@slac.stanford.edu}
\\
$^1$Mathematical Sciences and STAG Research Centre, University of Southampton,
Southampton, SO17 1BJ, UK
}
\date{Accepted 2015 June 11. Received 2015 June 9; in original form 2015 June 1}

\begin{document}
\pagerange{\pageref{firstpage}--\pageref{lastpage}} \pubyear{2015}

\maketitle

\label{firstpage}

\begin{abstract}
In the standard scenario for spin evolution of isolated neutron stars,
a young pulsar slows down with a surface magnetic field $B$ that does not
change.  Thus the pulsar follows a constant $B$ trajectory in the phase
space of spin period and spin period time derivative.
Such an evolution predicts a braking index $n=3$ while the field is constant
and $n>3$ when the field decays.
This contrasts with all nine observed values being $n<3$.
Here we consider a magnetic field that is buried soon after birth and
diffuses to the surface.
We use a model of a growing surface magnetic field to fit observations of
the three pulsars with lowest $n$: PSR~J0537$-$6910 with $n=-1.5$,
PSR~B0833$-$45 (Vela) with $n=1.4$, and PSR~J1734$-$3333 with $n=0.9$.
By matching the age of each pulsar, we determine their magnetic field
and spin period at birth
and confirm the magnetar-strength field of PSR~J1734$-$3333.
Our results indicate that all three pulsars formed in a similar way to central
compact objects (CCOs), with differences due to the amount of accreted mass.
We suggest that magnetic field emergence may play a role in the distinctive
glitch behaviour of low braking index pulsars, and we propose glitch behaviour
and characteristic age as possible criteria in searches for CCO descendants.
\end{abstract}

\begin{keywords}
stars: magnetic field
-- stars: neutron
-- pulsars: general
-- pulsars: individual: PSR~J0537$-$6910
-- pulsars: individual: PSR~B0833$-$45
-- pulsars: individual: PSR~J1734$-$3333
\end{keywords}

\maketitle

\section{Introduction} \label{sec:intro}

\begin{figure*}
\begin{center}
\includegraphics[scale=0.6]{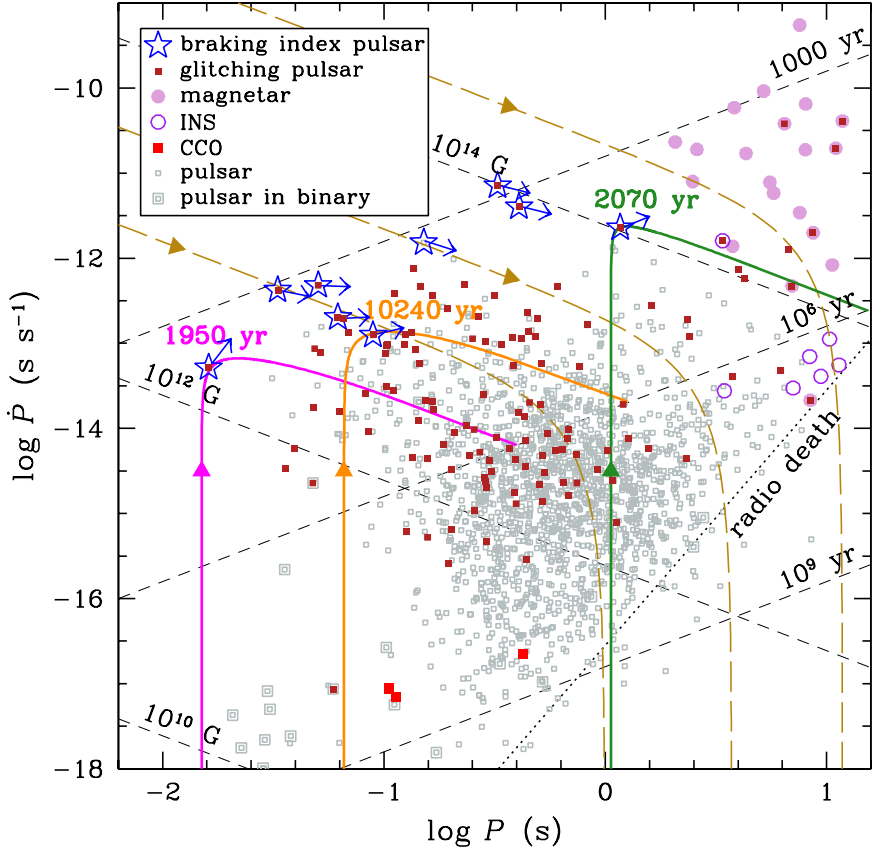}
\caption{
Pulsar spin period $P$ versus spin period time derivative $\Pdot$.
Open squares denote pulsars whose values are taken from the ATNF Pulsar
Catalogue; pulsars in a binary system and pulsars
that have been observed to glitch are also noted.
The nine stars denote pulsars with a measured braking index
(see Table~\ref{tab:psr}),
with arrows indicating the direction each pulsar is evolving,
as determined by their braking index.
Closed and open circles and closed squares denote magnetars and isolated
neutron stars (INSs) and central compact objects (CCOs), respectively.
Short-dashed lines indicate characteristic age $\tau_{\rm c}$ ($=P/2\Pdot$)
and inferred magnetic field $B$ [$=6.4\times10^{19}\mbox{ G }(P\Pdot)^{1/2}$].
Dotted line indicates the (theoretically uncertain) death line for pulsar
radio emission.
Solid lines indicate evolution trajectories (to age = $10^6\mbox{ yr}$)
where the magnetic field is buried initially and
(from left to right) the initial field strength and spin period and burial
density are [$B_0$(G),$P_0$(s),$\log\rho_{\rm b}(\mbox{g cm$^{-3}$})$] =
($3.4\times 10^{12}$,0.015,11.0), ($1.1\times 10^{13}$,0.066,11.5),
and ($1.3\times 10^{14}$,1.06,10.5), respectively.
Long-dashed lines indicate trajectories where the magnetic field evolves
according to eq.~(\ref{eq:magevol}) and
(from left to right) the initial field strength and field decay timescale are
[$B_0$(G),$\tau$(yr)] = ($8\times 10^{12}$,$10^6$), ($3\times 10^{13}$,$10^6$),
and ($3\times 10^{14}$,$10^5$), respectively.
}
\label{fig:ppdot}
\end{center}
\vspace{-0.2cm}
\end{figure*}

The magnetic field strength on the surface of neutron stars (NSs) spans
a wide range:
from $B\sim 10^8-10^9\mbox{ G}$ for millisecond pulsars and NSs in
low-mass X-ray binaries, through $10^{12}-10^{13}\mbox{ G}$ for normal
radio pulsars, to $10^{14}-10^{15}\mbox{ G}$ for magnetars.
The primary method used to determine these magnetic fields is by measuring
each pulsar's spin period $P$ and spin period time derivative $\dot{P}$.
Then assuming that the pulsar rotational energy decreases as a result
of emission of magnetic dipole radiation, the surface field strength $B$
at the magnetic
pole\footnote{Note that a coefficient of 3.2 in eq.~(\ref{eq:pdot}) is often
used in the literature, so that the inferred field in such a case is the field
at the magnetic equator.  Since we model field evolution at the magnetic pole,
hereafter we only refer to the value at the pole.}
is inferred, i.e.,
\be
P\dot{P} = \frac{\gamma}{2}B^2
\quad \Leftrightarrow \quad
B = 6.4\times 10^{19}\mbox{ G }(P\dot{P})^{1/2},
\label{eq:pdot}
\ee
where $\gamma=4\pi^2R^6\sin^2\alpha/3c^3I
= 4.884\times 10^{-40}\mbox{ s G$^{-2}$ }R_6^6I_{45}^{-1}\sin^2\alpha$,
$R$ and $I$ are the NS radius and moment of inertia, respectively,
$\alpha$ is the angle between the stellar rotation and magnetic axes,
and $R_6=R/10^6\mbox{ cm}$ and $I_{45}=I/10^{45}\mbox{ g cm$^2$}$
(\citealt{gunnostriker69}; see also \citealt{spitkovsky06,contopoulosetal14}).
Figure~\ref{fig:ppdot} shows the measured pulsar spin period and spin period
derivative values taken from the ATNF Pulsar
Catalogue\footnote{http://www.atnf.csiro.au/research/pulsar/psrcat/}
\citep{manchesteretal05}.
Also shown is a set of parallel lines which indicates the inferred magnetic
field obtained from eq.~(\ref{eq:pdot}).
The other set of parallel lines indicates the pulsar spin-down or
characteristic age $\tau_{\rm c}=P/2\dot{P}$;
$\tau_{\rm c}$ is often used as a surrogate for the true age of a pulsar.

The standard scenario for the rotational evolution of a pulsar is that it
is born rapidly spinning (e.g., with an initial spin period $P_0$ in the
millisecond regime) and rapidly slowing or spinning down, i.e., large $\dot{P}$.
This would place a newborn pulsar in the top-left region of
Fig.~\ref{fig:ppdot}.
As it spins down, the pulsar moves for $\sim 10^5-10^6\mbox{ yr}$ along a
$P$--$\dot{P}$ path that tracks one of the short-dashed lines of constant
magnetic field, evolving towards the bottom-right.
This is because magnetic field diffusion and decay occurs on the Ohmic
timescale
\be
\tau_{\mathrm{Ohm}}=\frac{4\pi\sigma_{\rm c} L^2}{c^2}\sim
 4\times 10^5\mbox{ yr}\lp\frac{\sigma_{\rm c}}{10^{23}\mbox{ s$^{-1}$}}\rp
 \lp\frac{L}{1\mbox{ km}}\rp^2, \label{eq:tohm}
\ee
where $\sigma_{\rm c}$ is the electrical conductivity, $L$ is the
lengthscale over which decay occurs, and 1~km is the approximate size of
the stellar crust (see Fig.~\ref{fig:eos});
note that magnetic field changes can occur earlier for magnetars due to
Hall effects which operate on a timescale
$\tau_{\mathrm{Hall}}=(4\pi en_{\rm e} L^2)/(cB)\sim 2\times 10^5\mbox{ yr}\,
(\rho/10^{12}\mbox{ g cm$^{-3}$})(B/10^{14}\mbox{ G})^{-1}(L/1\mbox{ km})^2$,
where $e$ is electron charge, $n_{\rm e}$ is electron number density, and
$\rho$ is mass density
\citep{goldreichreisenegger92} (see also \citealt{glampedakisetal11}).
In this work, we are primarily concerned with normal pulsars with
$B\sim 10^{12}-10^{13}\mbox{ G}$ and ages $<\mbox{a few}\times 10^4\mbox{ yr}$.
Thus at times $t\ll\tau_{\mathrm{Ohm}}$, the magnetic field does not change,
and $\tau_{\rm c}$ may be an adequate estimate of pulsar age
(see Section~\ref{sec:discuss}).
However when $t\rightarrow\tau_{\mathrm{Ohm}}$, the magnetic field decreases,
causing the efficiency of dipole radiation to decrease and
$\dot{P}\rightarrow 0$ [see eq.~(\ref{eq:pdot})].
Simple examples of such evolutionary paths are shown by the long-dashed curves
in Fig.~\ref{fig:ppdot}, for different initial magnetic field $B_0$
and decay timescale.
In particular, we assume a magnetic field that evolves with time as
\be
B(t) = \frac{B_0}{1+t/\tau}, \label{eq:magevol}
\ee
where $\tau$ is the field decay timescale which can be taken to be
approximately equal to $\tau_{\mathrm{Ohm}}$
(or $\tau_{\mathrm{Hall}}$ for magnetars).
Equation~(\ref{eq:magevol}) mimics the results of numerical simulations of
magnetic field evolution in the crust (see, e.g., \citealt{colpietal00}).
Observations and theoretical work seem to support the above scenario of an
approximately constant magnetic field early in the life of a pulsar
and a slowly decaying field at later times
(e.g., \citealt{viganoetal13} find that the magnetic field is constant
until an age of $\mbox{a few}\times 10^6\mbox{ yr}$ for
$B_0\lesssim 10^{14}\mbox{ G}$ and $\sim 10^5\mbox{ yr}$ for
$B_0\gtrsim 10^{14}\mbox{ G}$,
and \citealt{igoshevpopov14} find a field decay timescale of
$\sim 4\times 10^{5}\mbox{ yr}$).

On the other hand, there also exist observations that suggest that the magnetic
field evolves, and especially fields that grow, in young NSs.
An important example comes from the (measured) braking index of pulsars.
The second time derivative of the period $\ddot{P}$ can be determined in a
few pulsars (where $\ddot{P}$ is not dominated by timing noise),
and $\ddot{P}$ is conventionally expressed in terms of the braking index
$n$, which is given by
\be
n = 2-\frac{P\ddot{P}}{\dot{P}^2}. \label{eq:nbrake}
\ee
If pulsar spin-down is due to only magnetic dipole radiation and the field
is constant, then eq.~(\ref{eq:pdot}) yields a braking index $n=3$.
However, $n<3$ for all pulsars with a measured $\ddot{P}$
(see Table~\ref{tab:psr}).
The low observed values of $n$ can be attributed to a magnetic field that is
increasing: allowing $B$ to evolve in eq.~(\ref{eq:pdot}), one easily obtains
\be
n = 3 - 2\frac{\dot{B}}{B}\frac{P}{\dot{P}}
 = 3 - 4\tau_{\rm c}\frac{\dot{B}}{B}
 = 3 - \frac{4}{\gamma}\frac{\dot{B}}{B^3}P^2, \label{eq:nbrake2}
\ee
where $\dot{B}$ is time derivative of $B$.
For each pulsar with a measured braking index, we denote its possibly
evolving field in Fig.~\ref{fig:ppdot} by arrows directed along a pulsar's
trajectory in $P$--$\dot{P}$ phase space
(see also \citealt{espinozaetal11b,espinoza13}).
It is clear that about five of the nine pulsars are moving along trajectories
almost parallel to (short-dashed) lines of constant $B$, and these are pulsars
with braking index between 2 and 3.
Two pulsars (PSR~J0537$-$6910 and J1734$-$3333) are clearly crossing
lines of constant $B$, thus suggesting that their fields are growing and
J1734$-$3333 is evolving into a magnetar \citep{espinozaetal11b}.
Note that we assume a constant $\alpha$ for simplicity
[see eq.~(\ref{eq:pdot})].
An evolving $\alpha$ can produce similar spin evolution behaviour to one with
$\dot{B}$.  However evidence for a varying $\alpha$ is uncertain
(see, e.g., \citealt{lyneetal13,lyneetal15}), and for example,
\cite{guillonetal14} find that either $\alpha$ is constant or the timescale for
its variation is very long.

\begin{table*}
\begin{minipage}{170mm}
\caption{Pulsars with a measured braking index $n$.
Spin periods $P$ and period derivatives $\dot{P}$ are taken from the
ATNF Pulsar Catalogue.
Number in parentheses is uncertainty in last digit of braking index.
Glitch data are taken from the Glitch Catalogue and from \citet{hoetal15},
and references therein.  References for age and braking index:
[1] \citealt{lyneetal93},
[2] \citealt{wanggotthelf98,chenetal06},
[3] \citealt{middleditchetal06},
[4] \citealt{parketal10},
[5] \citealt{gradarietal11},
[6] \citealt{pageetal09,tsurutaetal09},
[7] \citealt{lyneetal96},
[8] \citealt{kumaretal12},
[9] \citealt{weltevredeetal11},
[10] \citealt{gaensleretal99},
[11] \citealt{livingstonekaspi11},
[12] \citealt{hoandersson12},
[13] \citealt{espinozaetal11b},
[14] \citealt{bocchinoetal05,camiloetal06},
[15] \citealt{royetal12},
[16] \citealt{blantonhelfand96},
[17] \citealt{livingstoneetal07,livingstoneetal11}.
}
\label{tab:psr}
\begin{tabular}{lcccccccc}
\hline
Pulsar & SNR & $P$ & $\dot{P}$ & $\tau_{\rm c}$ & Age & Braking & No. of
 & Typical \\
& & (s) & (s s$^{-1}$) & (yr) & (yr) & index $n$ & glitches
 & $\Delta\Omega/\Omega$ \\
\hline
B0531$+$21 & Crab & 0.0331 & 4.23$\times 10^{-13}$ & 1240 & 961 & 2.51(1) [1] & 25 & $10^{-9}-10^{-8}$ \\
J0537$-$6910 & N157B & 0.0161 & 5.18$\times 10^{-14}$ & 4930 & 2000$^{+3000}_{-1000}$ [2] & $-1.5$(1) [3] & 45 & $10^{-7}$ \\
B0540$-$69 & 0540$-$69.3 & 0.0505 & 4.79$\times 10^{-13}$ & 1670 & 1000$^{+660}_{-240}$ [4] & 2.087(7) [5] & 1 & $10^{-9}$ \\
B0833$-$45 & Vela & 0.0893 & 1.25$\times 10^{-13}$ & 11300 & 11000$^{+5000}_{-5600}$ [6] & 1.4(2) [7] & 19 & $10^{-6}$ \\
J1119$-$6127 & G292.2$-$0.5 & 0.408 & 4.02$\times 10^{-12}$ & 1610 & 7100$^{+500}_{-2900}$ [8] & 2.684(2) [9] & 3 & $10^{-9}-10^{-6}$ \\
B1509$-$58 & G320.4$-$1.2 & 0.151 & 1.53$\times 10^{-12}$ & 1570 & $<21000$ [10] & 2.832(3) [11] & 0 & --- \\
J1734$-$3333 & G354.8$-$0.8 & 1.17 & 2.28$\times 10^{-12}$ & 8130 & $>1300$ [12] & 0.9(2) [13] & 1 & $10^{-7}$ \\
J1833$-$1034 & G21.5$-$0.9 & 0.0619 & 2.02$\times 10^{-13}$ & 4850 & 1000$^{+200}_{-800}$ [14] & 1.8569(6) [15] & 4 & $10^{-9}$ \\
J1846$-$0258 & Kesteven 75 & 0.327 & 7.11$\times 10^{-12}$ & 728 & 1000$^{+3300}_{-100}$ [16] & 2.65(1) [17] & 2 & $10^{-9}-10^{-6}$ \\
\hline
\end{tabular}
\end{minipage}
\end{table*}

In this work, we describe an alternative to the standard scenario for pulsar
spin evolution described above, one that provides a physical mechanism for
a growing magnetic field, using the model of \cite{ho11}
(see also \citealt{muslimovpage96,geppertetal99,viganopons12}).
In brief, pulsars are born with a strong field, but this field was buried by
an early episode of accretion and is slowly diffusing to the surface.
In such cases, the surface field responsible for spin evolution
[including $n<3$ from eq.~(\ref{eq:nbrake2})]
is increasing at the current epoch.
Since glitches and timing noise could be responsible for small changes in
$n$ (see \citealt{livingstoneetal11,antonopoulouetal15,lyneetal15}),
we only consider pulsars with a braking index less than half
that predicted by the standard scenario (i.e., $n<1.5$) as sources whose
braking index requires an explanation beyond the standard scenario
(see \citealt{espinozaetal11b,lyneetal15}, and references therein, for
discussion of other models for low braking index;
see also \citealt{hamiletal15}).
From Table~\ref{tab:psr} we see that this criterion is satisfied by
three pulsars: PSR~J0537$-$6910, B0833$-$45 (Vela), and J1734$-$3333;
note that \cite{muslimovpage96} considered B0531$+$21 (Crab), B0540$-$69,
and B1509$-$58 which have $n>2.1$.
In Section~\ref{sec:model}, we briefly describe the model for magnetic
field evolution.
In Section~\ref{sec:results}, we present our results and use the
measured braking index and age of the three pulsars to determine the
initial magnetic field and spin period of each pulsar and the amount
of matter each accreted.
In Section~\ref{sec:discuss}, we summarize our findings and discuss
their implications.

\vspace{-0.2cm}
\section{Magnetic field evolution model} \label{sec:model}

Our calculation of magnetic field evolution follows that of
\cite{urpinmuslimov92} and \cite{ho11}.
Here we provide a summary and describe updates
(see \citealt{ho11}, for details).
The magnetic field is assumed to be buried deep beneath the surface by a
post-supernova episode of hypercritical accretion
\citep{chevalier89,geppertetal99,bernaletal10,bernaletal13}.
The field then diffuses to the surface on a timescale that depends on
burial depth.
One might expect an anti-correlation between field growth rate and pulsar
velocity since a smaller amount of mass will be accreted if the pulsar is
moving at a greater velocity; observations tentatively support such a
relation \citep{guneydaseksi13}.

To determine the evolution of the buried magnetic field, we solve the
induction equation
\be
\frac{\partial{\mathbf B}}{\partial t}
 =-\nabla\times\lp\frac{c^2}{4\pi\sigma_{\rm c}}\nabla\times{\mathbf B}\rp.
\label{eq:induction}
\ee
Our interest is in the NS crust, which is predominantly in a solid
state, and thus we neglect internal fluid motion.
We take the surface field after NS formation, but prior to mass accretion,
as the magnetic field strength at birth $B_0$.
Accretion then buries and compresses this birth field.
We assume a dipolar field in the stellar interior.
In \cite{ho11}, we consider two field configurations: one in which the
field is confined in the crust and another in which the field extends
into the core.
In the crust-confined case, the surface field grows at first but then decays
at later times.  In the crust-core field case, the surface field
strength grows until it saturates at the level of the core field, and
the core field decays on a much longer timescale ($>10^6\mbox{ yr}$,
especially if the core is superconducting).
Results for these two cases are qualitatively similar during the epoch
of field growth.
Here we consider only the latter case and hold the core field strength constant
(see \citealt{luoetal15}, for more results with a crust-confined field).
A constant core field is justified since the Ohmic decay timescale
[see eq.~(\ref{eq:tohm})] in the core is longer than our times of
interest ($t\ll 10^{6}\mbox{ yr}$).

\begin{figure}
\begin{center}
\includegraphics[scale=0.42]{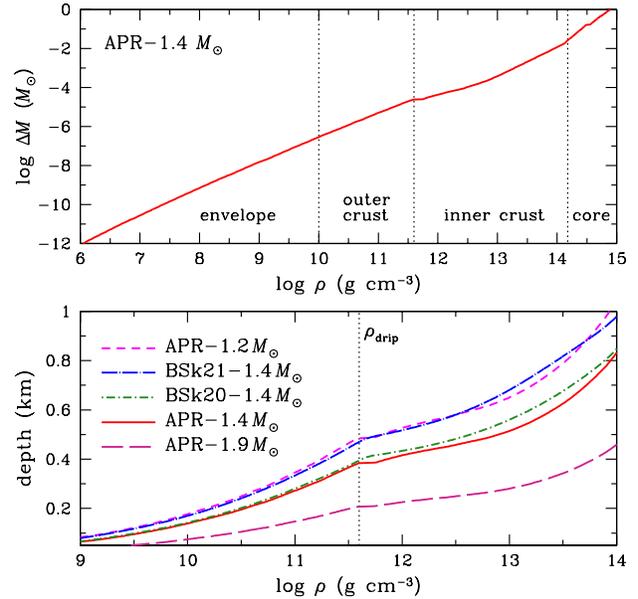}
\caption{
Accreted mass $\Delta M$ (top panel) and depth (bottom panel) as a function
of density $\rho$.
Vertical dotted lines indicate boundaries between
envelope and crust (taken to be at $10^{10}\mbox{ g cm$^{-3}$}$),
outer and inner crust (at
$\rho_{\rm drip}\approx 4\times 10^{11}\mbox{ g cm$^{-3}$}$),
and crust and core (at baryon density $n=0.09\mbox{ fm$^{-3}$}$).
}
\label{fig:eos}
\end{center}
\vspace{-0.2cm}
\end{figure}

We updated the magnetic field evolution code used in \cite{ho11} in
primarily two ways.
First, we use NS crust and core models built with the APR \citep{akmaletal98},
BSk20, or BSk21 \citep{potekhinetal13} models of the nuclear equation of
state (EOS).
In the top panel of Fig.~\ref{fig:eos}, we show the mass above a given
density $\Delta M$ [$\equiv M-m(r)$, where $M$ is total NS mass and
$m(r)$ is mass enclosed within radius $r$]. $\Delta M$ is an indication
of the amount of accreted mass needed to bury the magnetic field to a
given density.
The bottom panel shows the crust depth ($\equiv R-r$) as a function of
density for different NS masses and EOSs.
While $\Delta M(\rho)$ is weakly dependent on total NS mass and EOS,
depth versus density is a strong function of $M$ and EOS.
For a given EOS model (e.g., APR), crust thickness decreases with increasing
mass.  Therefore to bury the magnetic field at a particular density, the
field must be buried at a greater depth for a lower $M$.
For different EOS models, we see that the result for BSk20 is very similar
to that for APR, and thus the field evolution timescale [which scales with
depth, or $L$, as given by eq.~(\ref{eq:tohm}) for the Ohmic timescale]
for BSk20 and APR will be comparable.
Depth at a given density for NSs built using BSk21 is larger than that for
NSs built using APR, and thus the field evolution timescale for BSk21 will
be longer than that for APR.
Note that, in addition to depth variations between the three EOS models
considered here, crust composition for each model is different,
which in turn produces different electrical conductivities.
We mention that the previous works of \cite{muslimovpage96} and
\cite{geppertetal99} consider different EOS models than those studied here.

The second update is that we use
{\small CONDUCT13}\footnote{http://www.ioffe.ru/astro/conduct/},
which implements the latest advancements in calculating electrical
conductivities \citep{potekhinetal15}.
We assume no contribution due to impurity scattering since this only
becomes important at high densities and low temperatures.
We checked that there are no noticeable changes for a uniform impurity
parameter $Q_{\rm imp}\le 1$ (where $\sigma_{\rm c}\propto Q_{\rm imp}^{-1}$),
which is the relevant regime for the crust
of isolated NSs, in contrast to that of NSs accreting from a binary companion.
Recent works examine the effects of larger $Q_{\rm imp}$ on spin and
magnetic field evolution \citep{ponsetal13,viganoetal13,horowitzetal15}.
However, this occurs due to pasta phases near the crust-core boundary at
densities $\sim 10^{14}\mbox{ g cm$^{-3}$}$, and its effects only become
important after $\sim 10^5\mbox{ yr}$ \citep{ponsetal13,viganoetal13}.

\vspace{-0.2cm}
\section{Fit to low braking index pulsars}
\label{sec:results}

The age of the three pulsars (PSR~J0537$-$6910, B0833$-$45, and J1734$-$3333)
with $n<1.5$ is in the range of $\approx 1$--16~kyr (see Table~\ref{tab:psr}).
Therefore, we seek a magnetic field growth timescale on this order.
From eq.~(\ref{eq:tohm}) and Fig.~\ref{fig:eos}, we estimate that the
magnetic field should be buried at a density $\sim 10^{11}\mbox{ g cm$^{-3}$}$,
which corresponds to an accreted mass $\Delta M\sim 10^{-5}M_\odot$.

Beginning with an initial condition which has the magnetic field $B_0$
buried at density $\rho_{\rm b}$, the induction equation
[eq.~(\ref{eq:induction})] is solved to obtain the radial profile of
magnetic field as a function of time.
Integrating eq.~(\ref{eq:pdot}), we obtain the evolution of the spin period
\be
P(t) = \left[ P_0^2+\gamma\int_{t_0}^tB(R,t')^2\,dt'\right]^{1/2},
\ee
where $P_0$ is the initial pulsar spin period and $B(R,t)$ is the magnetic
field at the NS surface at time $t$ and is given by our solution to the
induction equation.
From $P(t)$, we calculate its first and second time derivatives
$\dot{P}(t)$ and $\ddot{P}(t)$, respectively, and
corresponding braking index $n$ given by eq.~(\ref{eq:nbrake2}).
For each of the three pulsars, we vary the three initial conditions
[$B_0$(G),$P_0$(s),$\log\rho_{\rm b}(\mbox{g cm$^{-3}$})$]
until the resulting $P$, $\dot{P}$, and $n$ match those of the pulsar at its
current age.
It is worth pointing out (see also \citealt{ho13b}) that fitting to observables
$P$ and $\dot{P}$ is equivalent to fitting to $\tau_{\rm c}$,
$B$ as given by eq.~(\ref{eq:pdot}), or spin-down luminosity since these
are all formed from $P$ and $\dot{P}$ (see, e.g., \citealt{muslimovpage96}),
while $n$ is an independent parameter since it includes $\ddot{P}$.

Figure~\ref{fig:ppdot} shows results for a set of initial conditions that
fit each of the three pulsars: ($B_0$,$P_0$,$\log\rho_{\rm b}$) =
($3.4\times 10^{12}$,0.015,11.0) for J0537$-$6910,
($1.1\times 10^{13}$,0.066,11.5) for B0833$-$45,
and ($1.3\times 10^{14}$,1.06,10.5) for J1734$-$3333.
Each trajectory is labelled with the approximate age at which the calculated
$P$, $\dot{P}$, and $n$ match their corresponding observed values.
For J0537$-$6910, we obtain an age of 1.95~kyr, compared to its known age
of $2_{-1}^{+3}\mbox{ kyr}$.
Because of its young age, the current spin
period $P=16\mbox{ ms}$ is not much different from the initial spin period
$P_0=15\mbox{ ms}$.
For B0833$-$45, we obtain an age of 10.2~kyr, compared to its known age
of $11_{-5.6}^{+5}\mbox{ kyr}$.
The spin period in this pulsar has increased significantly from 66~ms to
the current 89~ms due to its much older age.
For J1734$-$3333, we obtain an age of 2.07~kyr, compared to its minimum age
of $>1.3\mbox{ kyr}$.
Despite its similar age to J0537$-$6910, J1734$-$3333 underwent noticeable
spin-down due to its much stronger magnetic field.
Since $B\sim 10^{14}\mbox{ G}$ for J1734$-$3333, Hall effects
and anisotropic conductivities may need to be taken into account,
although these should not change our conclusions at a qualitative level.
We also note the work of \cite{gourgouliatoscumming15}, who use numerical
simulations to model Hall drift effects and obtain field growth with
multipolar toroidal fields of the order of $10^{14}\mbox{ G}$ that can produce
braking indices matching those observed.

\begin{figure*}
\begin{center}
\resizebox{!}{0.32\textwidth}{
\includegraphics{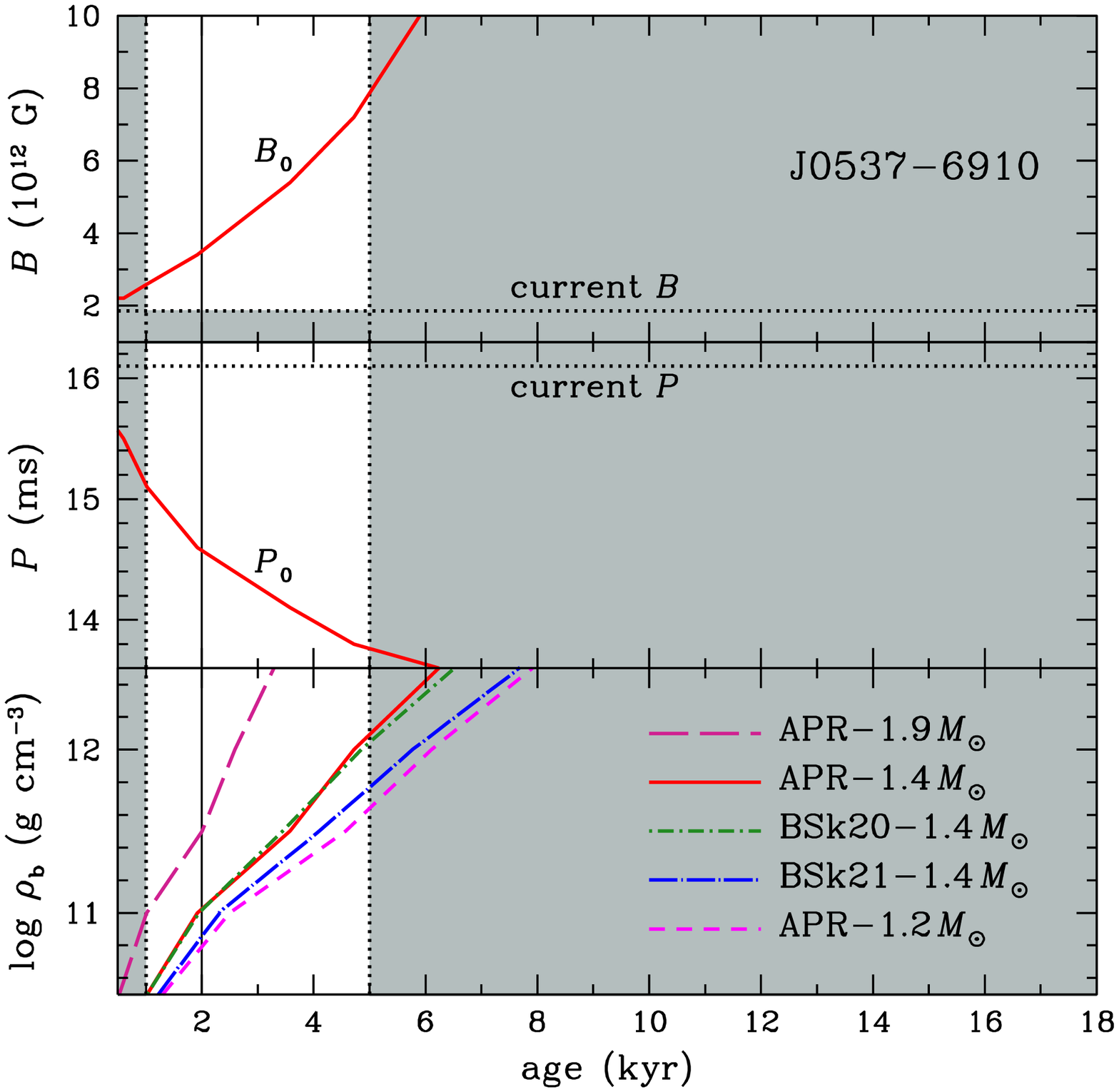} \includegraphics{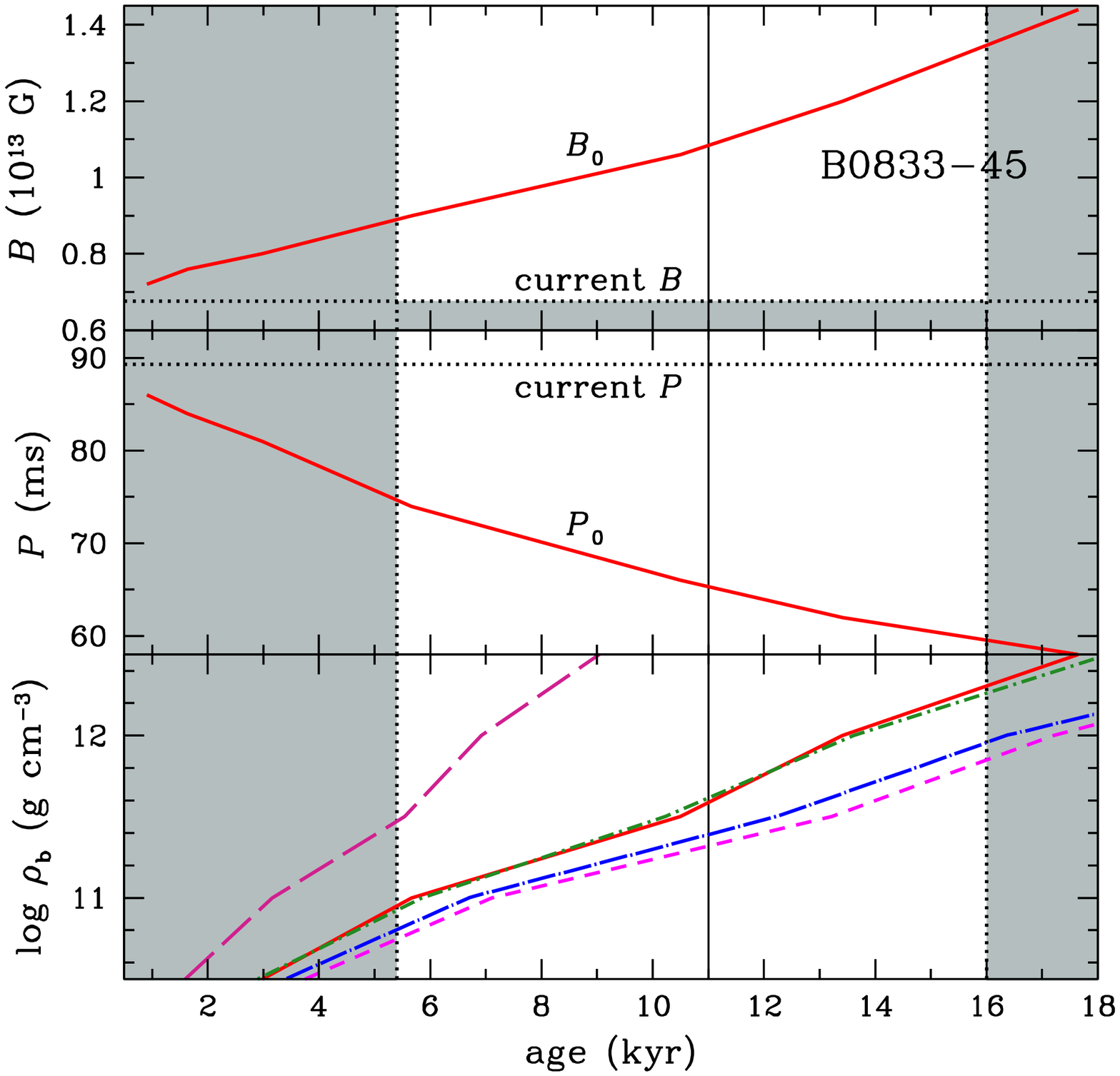} \includegraphics{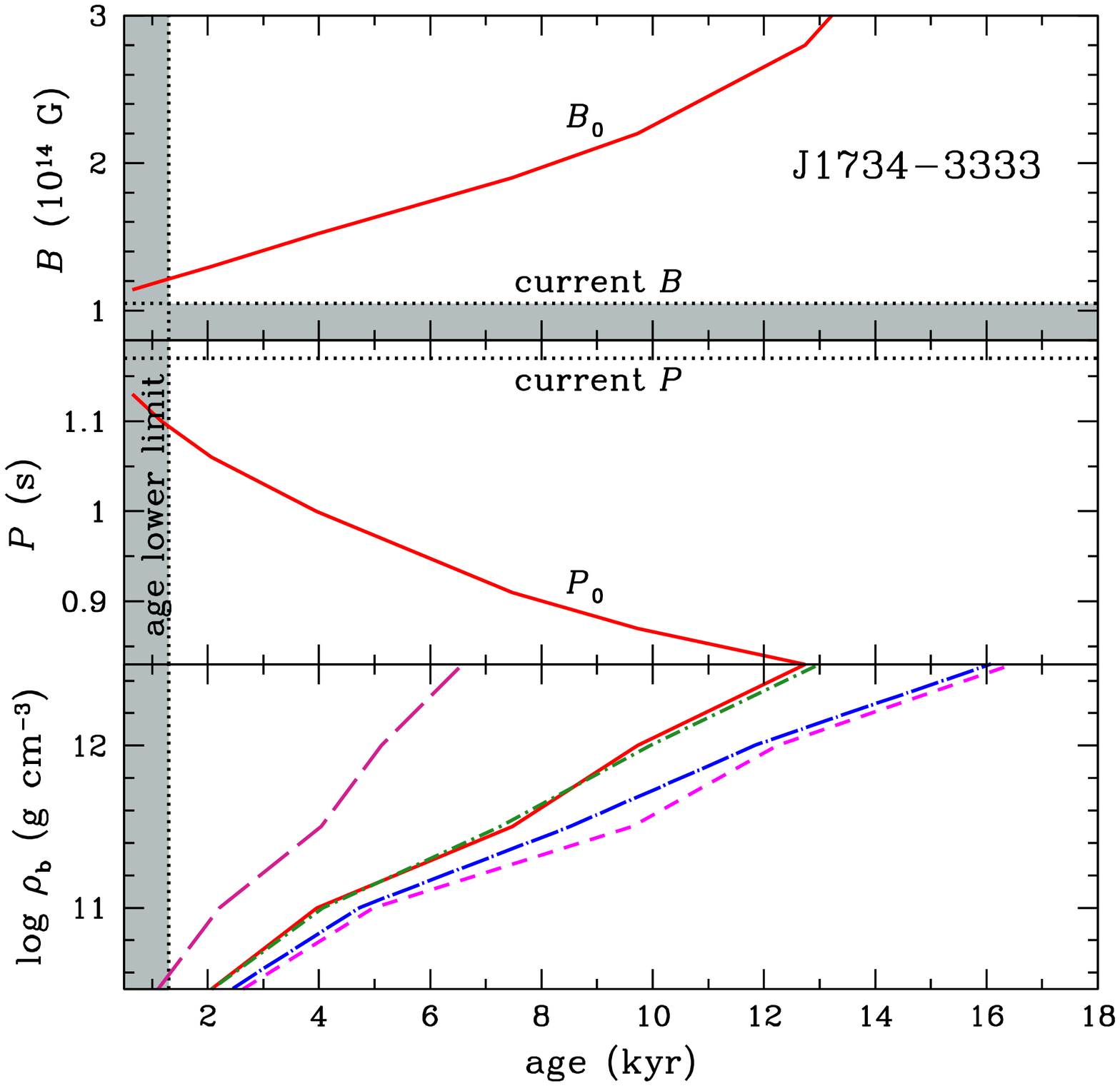}
}
\caption{
Initial magnetic field $B_0$ (top), initial spin period $P_0$ (centre),
and burial density $\rho_{\rm b}$ (bottom) as a function of age for
PSR~J0537$-$6910 (left-hand panel), B0833$-$45 (centre panel), and J1734$-$3333
(right-hand panel).
Horizontal dotted lines denote the current magnetic field $B$ and
spin period $P$ of each pulsar, where the former is given by
$B=6.4\times10^{19}\mbox{ G }(P\Pdot)^{1/2}$.
Vertical solid and dotted lines indicate the nominal age and age range,
respectively, of each pulsar (see Table~\ref{tab:psr}),
except in the case of J1734$-$3333, the vertical dotted lines indicate the
age lower limit.
}
\label{fig:bpd}
\end{center}
\vspace{-0.4cm}
\end{figure*}

Figure~\ref{fig:bpd} shows relationships between the three initial conditions,
$B_0$, $P_0$, and $\rho_{\rm b}$,
as well as how initial values compare to present values.
For a pulsar at a given age, the inferred density at which the magnetic
field is initially buried depends on NS mass and nuclear EOS model, while
$B_0$ and $P_0$ do not depend significantly on $M$ or EOS.
The thicker the crust (see Fig.~\ref{fig:eos}), the shallower the field
needs to be buried to match current values of $P$, $\dot{P}$, and $n$.
The shallower the field is buried, the closer the current magnetic field
and spin period are to their initial values.
Note that $B$ is evolving towards $B_0$, while $P$ is evolving away from $P_0$.

More accurate age determinations would lead to tighter limits on the
burial density (and hence mass accreted, via Fig.~\ref{fig:eos}) and initial
magnetic field and spin period.
Current age estimates allow us to constrain the magnetic field strength
at birth to within a factor of about three, i.e.,
$B_0\approx (3-7)\times 10^{12}\mbox{ G}$ for J0537$-$6910,
$B_0\approx (0.9-1.3)\times 10^{13}\mbox{ G}$ for B0833$-$45,
and $B_0\sim (1-3)\times 10^{14}\mbox{ G}$ for J1734$-$3333.
These field strengths fall within the lognormal distributions determined
from population synthesis studies, which find an average and width $\sigma$
of $\log B_0=12.95\pm 0.55$ in the case where there is no field decay
\citep{fauchergiguerekaspi06,guillonetal14}
and $\log B_0=13.25\pm 0.6$ in the case of where there is (model-dependent)
field decay \citep{popovetal10,guillonetal14}.
The initial spin periods are somewhat shorter for J0537$-$6910
and B0833$-$45 and much longer for J1734$-$3333 than the average
found in these population synthesis studies, i.e.,
$P_0=0.30\pm0.15\mbox{ s}$ \citep{fauchergiguerekaspi06}
and $P_0=0.25\pm0.1\mbox{ s}$ \citep{popovetal10}.
However \cite{guillonetal14} find a much broader distribution, such that
even $P_0\approx 1\mbox{ s}$ for J1734$-$3333 is within $2\sigma$ of the
average value (see also \citealt{igoshevpopov13}).

\vspace{-0.2cm}
\section{Discussion} \label{sec:discuss}

The standard theoretical scenario for spin evolution of isolated
pulsars is one in which a pulsar loses its rotational energy via dipole
radiation and slows down over time with a constant surface magnetic
field, and this scenario works well to explain pulsars at an early age
($<10^5\mbox{ yr}$) and pulsars with braking index $n\approx 3$
(see Table~\ref{tab:psr}).
Here we study an alternative scenario, one in which the intrinsic magnetic
field of a pulsar is buried by accretion soon after its birth in a supernova.
Our scenario works alongside the standard scenario, by providing a natural
explanation for pulsars with $n<3$.
We fit observations of the three pulsars whose braking index $n<1.5$,
i.e., PSR~J0537$-$6910, B0833$-$45, and J1734$-$3333.
We find that the mass required to bury the magnetic field is
$\Delta M\sim 10^{-5}M_\odot$.
Another requirement is that the timescale over which this mass is accreted
must be shorter than the field diffusion timescale \citep{geppertetal99},
given approximately by eq.~(\ref{eq:tohm}).
Thus most newborn NSs probably fall within the standard scenario, with very
little or no accretion.
A relatively few, like the three examined here, quickly accreted enough mass
to bury their magnetic field.
Generally, the range of accreted mass could be quite large.  Thus a third
formation channel is one in which a large amount of matter is accreted.
Such is possibly the case for the central compact objects (CCOs) studied in
\cite{ho11}, where $\Delta M\gtrsim 10^{-4}M_\odot$.
CCOs have age $<\mbox{ a few}\times 10^4\mbox{ yr}$, and their surface
magnetic field is $\sim 10^{10}-10^{11}\mbox{ G}$
\citep{halperngotthelf10,gotthelfetal13,ho13}.

Unification of different observational classes of NSs, such as magnetars,
CCOs, and normal radio pulsars, via evolution of a NS from one class to
another \citep{kaspi10,popovetal10},
would help alleviate the NS-supernova birthrate problem \citep{keanekramer08}.
One pulsar studied here that is of particular interest in this regard is
J1734$-$3333.
Based on its current inferred magnetic field and braking index, this pulsar
appears to be moving in Fig.~\ref{fig:ppdot} from the region populated
by normal radio pulsars into that of magnetars \citep{espinozaetal11b}.
We find that it has a magnetar-strength magnetic field at birth, i.e.,
$B_0\sim (1-3)\times 10^{14}\mbox{ G}$, with the exact value dependent on
the age of the pulsar (see right-hand panel of Fig.~\ref{fig:bpd}).  In the case
of shallow field burial and no significant field decay during its life thus
far, the magnetic field of J1734$-$3333 is nearly at its birth value and
will reach it at an age $\approx 10^4\mbox{ yr}$.
We point out that our projection of the trajectory of J1734$-$3333 is one
that involves a braking index which changes over time
(cf. \citealt{espinozaetal11b}).
In particular, a detectable change from $n=0.9$ to $1$, based on the
current uncertainty of 0.1, takes about 70~yr.
J0537$-$6910 is perhaps a more promising target for observing a braking
index change: a change from $n=-1.5$ to $-1.4$ takes about 20~yr.
For completeness, the braking index change for Vela from $n=1.4$ to 1.5
takes about 400--500 yr.
The difficulty lies with the much larger braking index uncertainty for
these three (low braking index) pulsars compared to that of other pulsars
(see Table~\ref{tab:psr}).
For example, evolution of braking index and third time derivative of
spin period is observed for Crab \citep{lyneetal15} and B1509$-$58
\citep{livingstonekaspi11}.

It is possibly noteworthy that two (J0537$-$6910 and Vela) of the
three pulsars with $n<1.5$ undergo regular, large-amplitude glitches in
their timing behaviour.
The third (J1734$-$3333) recently had a large glitch, as reported in the
Glitch Catalogue\footnote{http://www.jb.man.ac.uk/pulsar/glitches.html}
\citep{espinozaetal11}.
The other pulsars with $1.8<n<3$ either are not seen to glitch (B1509$-$58),
have small amplitude glitches (e.g., Crab), or have glitches whose amplitude
varies greatly (e.g., the magnetar J1846$-$0258) (see Table~\ref{tab:psr}).
The distinctiveness of J0537$-$6910 and Vela glitches is discussed by
\cite{espinozaetal11}, who also show that glitch size has a bimodal
distribution (see also \citealt{yuetal13}).
The possible connection between low braking index and regular, large glitches
in certain pulsars is noted by \cite{espinoza13}, who report another three
pulsars (with $n=-1,2,$ and $2$) that could be in this group.
These large amplitude spin-up glitches are of great interest since they
may be revealing properties of the neutron superfluid in the NS
\citep{baymetal69,andersonitoh75,alparetal84}.
The regularity of these similarly-sized glitches is thought to be the
result of the pulsar tapping and exhausting the entire angular momentum
reservoir of the superfluid in the NS inner crust
\citep{linketal99,anderssonetal12,chamel13,piekarewiczetal14,hoetal15,hookeretal15,steineretal15}.
Our simulations of magnetic field diffusion from the (inner and
outer) crust to the surface suggest that perhaps this motion could be
involved in triggering glitches of the type seen in low braking index pulsars.

If regular, large glitches are a symptom of a previously buried magnetic
field, then glitch activity could be used as a criterion for searches
for descendants of CCOs.
Previous searches for CCOs and their descendants focus on the region in
$P$--$\dot{P}$ phase space where known CCOs reside (see Fig.~\ref{fig:ppdot}).
These searches have thus far not found definitive candidates
\citep{gotthelfetal13b,bogdanovetal14,luoetal15}.
As we show here (see also \citealt{luoetal15}), pulsars with an emerging
magnetic field move rapidly through this region of $P$--$\dot{P}$, and thus
the likelihood of discovery here is low.
This can also explain the relative paucity of pulsars here
\citep{halperngotthelf10,kaspi10}.
Once their intrinsic fields reach the surface and these
pulsars evolve to join the majority of the NS population, they might still
be distinguished by their glitch activity.
After all, glitch size and activity peak at age $\sim 10^4\mbox{ yr}$
\citep{mckennalyne90,espinozaetal11},
and glitches only occur in pulsars with $\tau_{\rm c}\lesssim 10\mbox{ Myr}$
\citep{espinozaetal11}.

Finally, it is well known that pulsar characteristic age $\tau_{\rm c}$
is often discrepant with true age, and thus the former can be an unreliable
estimate of the latter (e.g., the case with CCOs).
For seven of the nine braking index pulsars, the characteristic age is within
a factor of about two of the true age (see Table~\ref{tab:psr}), although
some of the true age determinations are likely biased towards $\tau_{\rm c}$.
Thus $\tau_{\rm c}$ is a relatively good age estimate for these types
of sources and when these pulsars are near maximum $\dot{P}$ and
post-maximum.  Thus low $\tau_{\rm c}$ could be another possible criterion
in searches for CCO descendants.

\vspace{-0.6cm}
\section*{acknowledgments}
WCGH thanks the anonymous referee for helpful comments.
The author also appreciates use of computer facilities at the Kavli Institute
for Particle Astrophysics and Cosmology and
acknowledges support from the Science and Technology
Facilities Council (STFC) in the United Kingdom.

\vspace{-0.6cm}
\bibliographystyle{mnras}

\label{lastpage}

\end{document}